\documentclass[preprint, aps, showpacs]{revtex4}
\usepackage{bm}
\begin{document}

\title{Extended Quark Potential Model from Random Phase Approximation}

\author{Weizhen Deng} 
\author{Xiaolin Chen}
\author{Dahai Lu}
\author{Liming Yang}
\affiliation{Department of Physics, Peking University, Beijing 100871, China}

\begin{abstract}
The quark potential model is extended to include the sea quark
excitation using the random phase approximation (RPA).  The effective
quark interaction preserves the important Quantum Chromodynamics (QCD)
properties --- chiral symmetry and confinement simultaneously.  A
primary qualitive analysis shows that the $\pi$ meson as a well-known
typical Goldstone boson and the other mesons made up of valence
$q\bar{q}$ quark pair such as the $\rho$ meson can also be described
in this extended quark potential model.

\end{abstract}

\pacs{12.39.Pn, 14.40Aq}

\maketitle

\section{Introduction}

The Quantum Chromodynamics (QCD) is widely believed to be the correct
theory of strong interaction nowdays.  Due to the complication of
non-Abelian gauge theory, in the studies of hadron structure, models
are often adopted to approximate the strong interaction.  Two kinds of
models are often used, each incorporates several important QCD
features.  The quark potential model based on the confinment quark
picture has been impressively successful in hadron spectra
\cite{Isgur}, except for the $\pi$ meson.  Recently, it was shown that
for the $\pi$ meson, which is a well-known Goldstone boson, the
explanation of quark potential model, i.e. mesons merely made up of
$q\bar{q}$ pairs of valence quarks, is not well satisfied and the
important sea quark contribution must be considered \cite{Cotanch}.
On the other hand, the Nambu--Jona-Lasinio (NJL) model \cite{Nambu}
and several of its extensions \cite{Bernard84,Langfeld96} can well
describe the $\pi$ meson as a Goldstone boson which is a conclusion of
QCD chiral symmetry.  The sea quark contributions to the meson
structure are included in the NJL model by using the Bethe-Salpter
equation or its equivalence --- Random Phase Approximation (RPA).
Also ground states of other mesons such as $\rho$ which are made up of
simple valence $q\bar{q}$ quark pair can be described by the
extensions to the original NJL model. However, the NJL model lacks the
important QCD property of quark confinement, so the excitation states
of mesons cannot be accounted in the model, such as $\pi'$, $\rho'$,
$\rho''$, etc.

In this paper, we try to incorporate the important QCD chiral symmetry
into the quark potential model by comprising the sea quark
contribution. To incorporate the chiral symmetry, we start from an
effective quark hamiltonian taken from extended NJL models which is
chiral invariant.  To consider the sea quark contribution, we adopt
the RPA approximation.  This derives a hamiltonian with two channels,
i.e. the ordinary valence quark channel and a new sea quark channel,
which is an extension to the quark potential model.  This extended
quark potential model with RPA approximation demands a new potential
which couples the valence quark channel with the sea quark channel we
call coupling potential.  Then a preliminary numerical analysis is
made to illustrate that the Goldstone $\pi$ meson and also $\rho$
meson can be well described together in the extended quark potential
model.

In the next section, We obtain the extended quark potential model ---
the RPA approximation is used to derive the two channel hamiltonian
and the so called coupling potential.  In sec.~3, a preliminary
analysis is made to the $\pi$ and $\rho$ mesons.  Finally, some brief
summary and discussion are given.

\section{Extended Quark Potential Model with RPA}

In this section, we will incorporate the QCD chiral symmetry into the
original quark potential model. To avoid the complication of QCD, our
start point will be an effective quark interaction.  A good choice of
the effective quark interaction with chiral symmetry is of the NJL
type.  However, despite of much success of the original NJL model, its
contact quark interaction is too simplified as an effective quark
interaction which does not possess the important QCD feature of color
confinement.  The original NJL model has been modified by several
authors to include the nonlocal effective quark interaction
\cite{Bernard84,Langfeld96}.

Consider a quark nonlocal action where the QCD strong interaction is
approximated by the quark effective interaction after the complicated
degrees of freedom of gluon are eliminated as in the extended NJL
models. Its general form is
\begin{eqnarray}
\label{A1}
A & = & \int d^4x \bar\Psi(x) (\gamma \cdot p - m_0) \Psi (x) \nonumber \\  
  & + & \frac12 \sum_{\epsilon} \int d^4x d^4y K_\epsilon(x-y) 
         \bar\Psi(x) \Gamma_\epsilon \Psi(x)
	 \bar\Psi(y) \Gamma_\epsilon \Psi(y), 
\end{eqnarray}
where $K_\epsilon(x)$ are the interaction kernel functions and
$\Gamma_\epsilon$ are tensor products of Dirac, flavor (in this paper,
we will restrict to iso-spin $SU(2)$ symmetry only) and color
($SU(3)$) matrices which represent the quark interaction channels. The
quark interaction in this quark action will be chosen to have chiral
symmetry and with quark confinement in the next section.

As the action is nonlocal, the exact hamiltonian will be very
complicated and also time dependent.  In the present paper we shall
only consider the non-relativistic limit.  We can adopt a simple,
time independent hamiltonian with an instantaneous approximation.
After approximating the kernel function as
\begin{equation}
K_\epsilon(x-y) \approx K_\epsilon({\bm x} - {\bm y}) \delta(x^0 - y^0), 
\end{equation}
the model hamiltonian becomes
\begin{eqnarray}
\label{H}
H & = & H_2 + H_4, \\
H_2 &=& \int d^3x \Psi^\dagger({\bm x}) (-i {\bm\alpha} \cdot {\bm\nabla} 
         + \beta m_q) \Psi({\bm x}), \\  
H_4 &=& - \frac12 \sum_\epsilon \int d^3x d^3y K_\epsilon({\bm x} - {\bm y})
         \bar\Psi({\bm x}) \Gamma_\epsilon \Psi({\bm x})
	 \bar\Psi({\bm y}) \Gamma_\epsilon \Psi({\bm y}), 
\end{eqnarray}
where $\Psi({\bm x}) \equiv \Psi({\bm x},0)$ which can be expressed in
quark creation and annhilation operators
\begin{eqnarray}
\Psi({\bm x}) &=& \sum_s \int \frac{d^3k}{(2\pi)^3} \frac1{2k^0}
\left[b(\vec{k},s) u(\vec{k},s) e^{i {\bm k}\cdot{\bm x}}+
d^\dagger(\vec{k},s) v(\vec{k},s) e^{-i {\bm k}\cdot{\bm x}} \right].
\end{eqnarray}
The hamiltonian from the instantaneous approximation breaks the
important Lorentz invariance.  We shall only use it to study the
hadron structure in rest frame.  For a moving hadron, however, we
shall use the Lorentz boost technique to restore the Lorentz
invariance in future study.

Apart from a constant, the hamiltonian (\ref{H}) can be written in
normal order form as
\begin{equation}
\label{HP}
H = H_0 + :H_4:
\end{equation}
where $H_0$ represents the single particle energy of constituent
quark and it can be re-written formly as
\begin{equation}
H_0 = \int d^3x \Psi^\dagger({\bm x}) \left[ -i A(- {\bm\nabla}^2)
      {\bm\alpha} \cdot {\bm\nabla} + \beta B(- {\bm\nabla}^2) \right] 
      \Psi({\bm x}). 
\end{equation}
The function forms of $A$ and $B$ operators should be obtained from
the Mean Field Approximation (MFA) and their momentum dependence are
due to the Lorentz invariance breaking from instantaneous
approximation.  Here for simplification we shall take the $A$ and $B$
operators as constants.  This only introduce one parameter of the
constituent quark mass $m_c$. The constituent quark single energy is
simplified as
\begin{equation}
\label{H00}
H_0 = \int d^3x \Psi^\dagger({\bm x}) \left[ -i 
      {\bm\alpha} \cdot {\bm\nabla} + \beta m_c \right] 
      \Psi({\bm x}). 
\end{equation}

It is generally accepted that hadrons are excited states of the QCD
vacuum.  To account the sea quark contribution, we consider a meson as
a superposition of creation and annihilation of q-$\bar{\mbox{q}}$
pairs on the vacuum.  Thus the meson states are treated as an excitation
modes of the vacuum of the RPA type:
\begin{equation}
\left| Q \right\rangle = Q^\dagger \left| 0 \right\rangle,
\end{equation}
where $Q^\dagger$ is the excitation operator.  For a meson in the rest
frame, $Q^\dagger$ can be expressed in quark and antiquark creation
and annihilation operators
\begin{eqnarray}
\label{Qop}
Q^\dagger & = & \sum_{s_1 s_2} \int \frac{d^3k}{(2\pi )^3} \frac 1{2k^0}
 \left[ x({\bm k},s_1,s_2) 
 b^\dagger({\bm k} s_1) d^\dagger(-{\bm k} s_2) \right. \nonumber \\  
 &  & \left. + y({\bm k},s_1,s_2) 
 \widetilde{d}(-{\bm k} s_1) \widetilde{b}({\bm k} s_2) \right], 
\end{eqnarray}
where $x$ and $y$ are the x-amplitude and y-amplitude of the RPA wave
function respectively and the time reversal operators are
\begin{eqnarray}
\widetilde{b}({\bm k} s) &\equiv& (-1)^{1/2-s} b({\bm k} -s), \\
\widetilde{d}({\bm k} s) &\equiv& (-1)^{1/2-s} d({\bm k} -s).
\end{eqnarray}

The x- and y-amplitudes can be solved from the RPA equation of
motion
\begin{equation}
\label{EOM}
\left\langle 0 \right| \left[ \delta Q,[H,Q^\dagger] \right] 
\left| 0 \right\rangle 
 = E \left\langle 0 \right| [\delta Q,Q^\dagger]
\left| 0 \right\rangle. 
\end{equation}
After inserting the hamiltonian eq.~(\ref{HP}) into the RPA equation of
motion, we obtain the equations for the x- and y- amplitudes
\begin{eqnarray} 
&&\left(E - 2 k^0 \right) x({\bm k},s_1,s_2) \nonumber \\
& = & - \sum_{s_3s_4} \int \frac{d^3k^{\prime}}{(2\pi)^3} 
     \frac1{2k^0} \frac1{2k^{\prime 0}}
     \sum_{\epsilon} K_{\epsilon} (0) 
     \nonumber \\  
&  & \left\{ \bar{u}({\bm k}s_1) \Gamma_\epsilon v(-{\bm k}s_2)
             \bar{v}(-{\bm k}'s_4) \Gamma_\epsilon u({\bm k}'s_3)
	     x({\bm k}',s_3,s_4) \right. \nonumber \\ 
&  & \left. -\bar{u}({\bm k}s_1) \Gamma_\epsilon v(-{\bm k}s_2)
             \widetilde{\bar{u}}({\bm k}'s_4) \Gamma_\epsilon 
	     \widetilde{v}(-{\bm k}'s_3)
	     y({\bm k}',s_3,s_4) \right\} \nonumber \\ 
& + & \sum_{s_3s_4} \int \frac{d^3k^{\prime}}{(2\pi)^3} 
     \frac1{2k^0} \frac1{2k^{\prime 0}}
     \sum_{\epsilon} K_{\epsilon} ({\bm k}-{\bm k}^\prime) 
     \nonumber \\  
&  & \left\{ \bar{u}({\bm k}s_1) \Gamma_\epsilon u({\bm k}'s_3)
             \bar{v}(-{\bm k}'s_4) \Gamma_\epsilon v(-{\bm k}s_2)
	     x({\bm k}',s_3,s_4) \right. \nonumber \\ 
&  & \left. -\bar{u}({\bm k}s_1) \Gamma_\epsilon \widetilde{v}(-{\bm k}'s_3)
             \widetilde{\bar{u}}({\bm k}'s_4) \Gamma_\epsilon v(-{\bm k}s_2)
	     y({\bm k}',s_3,s_4) \right\}, \\
\noalign{and} 
&&\left(E + 2 k^0 \right) y({\bm k},s_1,s_2) \nonumber \\
& = & - \sum_{s_3s_4} \int \frac{d^3k^{\prime}}{(2\pi)^3} 
     \frac1{2k^0} \frac1{2k^{\prime 0}}
     \sum_{\epsilon} K_{\epsilon} (0) 
     \nonumber \\  
&  & \left\{ \widetilde{\bar{v}}(-{\bm k}s_1) \Gamma_\epsilon 
	     \widetilde{u}({\bm k}s_2)
             \bar{v}(-{\bm k}'s_4) \Gamma_\epsilon u({\bm k}'s_3)
	     x({\bm k}',s_3,s_4) \right. \nonumber \\ 
&  & \left. -\widetilde{\bar{v}}(-{\bm k}s_1) \Gamma_\epsilon 
	     \widetilde{u}({\bm k}s_2)
             \widetilde{\bar{u}}({\bm k}'s_4) \Gamma_\epsilon 
	     \widetilde{v}(-{\bm k}'s_3)
	     y({\bm k}',s_3,s_4) \right\} \nonumber \\ 
& + & \sum_{s_3s_4} \int \frac{d^3k^{\prime}}{(2\pi)^3} 
     \frac1{2k^0} \frac1{2k^{\prime 0}}
     \sum_{\epsilon} K_{\epsilon} ({\bm k}-{\bm k}^\prime) 
     \nonumber \\  
&  & \left\{ \widetilde{\bar{v}}(-{\bm k}s_1) \Gamma_\epsilon u({\bm k}'s_3)
             \bar{v}(-{\bm k}'s_4) \Gamma_\epsilon \widetilde{u}({\bm k}s_2)
	     x({\bm k}',s_3,s_4) \right. \nonumber \\ 
&  & \left. -\widetilde{\bar{v}}(-{\bm k}s_1) \Gamma_\epsilon 
	     \widetilde{v}(-{\bm k}'s_3)
             \widetilde{\bar{u}}({\bm k}'s_4) \Gamma_\epsilon 
	     \widetilde{u}({\bm k}s_2)
	     y({\bm k}',s_3,s_4) \right\},  
\end{eqnarray}
where $\widetilde{u}$ and $\widetilde{v}$ are the time reversal of
spinors $u$ and $v$ respectively.  In each equation, the first
summation (direct term) can be absorbed into the second summation
(exchange term) by Fierz rearrangement, then equations briefly become
\begin{eqnarray} 
&&\left(E - 2 k^0 \right) x({\bm k},s_1,s_2) \nonumber \\
& = & \sum_{s_3s_4} \int \frac{d^3k^{\prime}}{(2\pi)^3} 
     \frac1{2k^0} \frac1{2k^{\prime 0}}
     \sum_{\epsilon} K_{\epsilon} ({\bm k}-{\bm k}^\prime) 
     \nonumber \\  
&  & \left\{ \bar{u}({\bm k}s_1) \Gamma_\epsilon u({\bm k}'s_3)
             \bar{v}(-{\bm k}'s_4) \Gamma_\epsilon v(-{\bm k}s_2)
	     x({\bm k}',s_3,s_4) \right. \nonumber \\ 
&  & \left. -\bar{u}({\bm k}s_1) \Gamma_\epsilon \widetilde{v}(-{\bm k}'s_3)
             \widetilde{\bar{u}}({\bm k}'s_4) \Gamma_\epsilon v(-{\bm k}s_2)
	     y({\bm k}',s_3,s_4) \right\}, \\ 
\noalign{and}
&&\left(E + 2 k^0 \right) y({\bm k},s_1,s_2) \nonumber \\
& = & \sum_{s_3s_4} \int \frac{d^3k^{\prime}}{(2\pi)^3} 
     \frac1{2k^0} \frac1{2k^{\prime 0}}
     \sum_{\epsilon} K_{\epsilon} ({\bm k}-{\bm k}^\prime) 
     \nonumber \\  
&  & \left\{ \widetilde{\bar{v}}(-{\bm k}s_1) \Gamma_\epsilon u({\bm k}'s_3)
             \bar{v}(-{\bm k}'s_4) \Gamma_\epsilon \widetilde{u}({\bm k}s_2)
	     x({\bm k}',s_3,s_4) \right. \nonumber \\ 
&  & \left. -\widetilde{\bar{v}}(-{\bm k}s_1) \Gamma_\epsilon 
             \widetilde{v}(-{\bm k}'s_3)
             \widetilde{\bar{u}}({\bm k}'s_4) \Gamma_\epsilon 
             \widetilde{u}({\bm k}s_2)
	     y({\bm k}',s_3,s_4) \right\}.  
\end{eqnarray}

Now We introduce two effective potentials $U$ and $V$, such that the matrix
elements are
\begin{eqnarray}
\label{U}
\langle {\bm k},s_1,s_2 \mid U \mid {\bm k}',s_3,s_4 \rangle 
& \equiv & \frac1{2k^0} \frac1{2k^{\prime 0}}
     \sum_{\epsilon} K_{\epsilon} ({\bm k}-{\bm k}^\prime) 
     \nonumber \\  
&  & \bar{u}({\bm k}s_1) \Gamma_\epsilon u({\bm k}'s_3)
     \bar{v}(-{\bm k}'s_4) \Gamma_\epsilon v(-{\bm k}s_2), \\ 
\label{V}
\langle {\bm k},s_1,s_2 \mid V \mid {\bm k}',s_3,s_4 \rangle
& \equiv & \frac1{2k^0} \frac1{2k^{\prime 0}}
     \sum_{\epsilon} K_{\epsilon} ({\bm k}-{\bm k}^\prime) 
     \nonumber \\  
&  & \bar{u}({\bm k}s_1) \Gamma_\epsilon \widetilde{v}(-{\bm k}'s_3)
     \widetilde{\bar{u}}({\bm k}'s_4) \Gamma_\epsilon v(-{\bm k}s_2).
\end{eqnarray}

From the basic symmetries of parity, charge conjugation and time
reversal of quark interaction, and the definition eqs. (\ref{U}) and
(\ref{V}), one can obtain the hermicity of effective potentials $U$ and
$V$
\begin{eqnarray}
U^\dagger = U, \\
V^\dagger = V,
\end{eqnarray}
and also
\begin{eqnarray}
&&\langle {\bm k}, s_1, s_2 \mid U \mid {\bm k}', s_3, s_4 \rangle^* 
\nonumber \\
&=& 
(-1)^{1/2 - s_1}
(-1)^{1/2 - s_2}
(-1)^{1/2 - s_3}
(-1)^{1/2 - s_4} \nonumber \\
&&\langle {\bm k}, -s_2, -s_1 \mid U \mid {\bm k}', -s_4, -s_3 \rangle.
\end{eqnarray}
Finally, the RPA equations are written as
\begin{eqnarray} 
&&\left(E - 2 k^0 \right) x({\bm k},s_1,s_2) \nonumber \\
& = & \sum_{s_3s_4} \int \frac{d^3k^{\prime}}{(2\pi)^3} 
\left[ \langle {\bm k},s_1,s_2 \mid U \mid {\bm k}',s_3,s_4 \rangle 
	     x({\bm k}',s_3,s_4) \right. \nonumber \\ 
&  & \left. -\langle {\bm k},s_1,s_2 \mid V \mid {\bm k}',s_3,s_4 \rangle 
	     y({\bm k}',s_3,s_4) \right], 
\label{XRPA}\\ 
\noalign{and}
&&\left(E + 2 k^0 \right) Y({\bm k},s_1,s_2) \nonumber \\
& = & \sum_{s_3s_4} \int \frac{d^3k^{\prime}}{(2\pi)^3} 
\left[ \langle {\bm k},s_1,s_2 \mid V \mid {\bm k}',s_3,s_4 \rangle 
	     x({\bm k}',s_3,s_4) \right. \nonumber \\ 
&  & \left. -\langle {\bm k},s_1,s_2 \mid U \mid {\bm k}',s_3,s_4 \rangle 
	     y({\bm k}',s_3,s_4) \right].  
\label{YRPA}
\end{eqnarray}

In the above RPA equations, the potential $U$ interacts between the
same RPA amplitudes (x-amplitude in eq.~(\ref{XRPA}) and y-amplitude
in eq.~(\ref{YRPA})). The x-amplitude represents the contribution of
valence quarks, that is just the wave-function of valence quarks in
the potential model. We conclude that the $U$ potential is just the
quark potential in the potential model. On the other hand, potential
$V$ interacts between the different RPA amplitudes. This is a new
potential and we call it the coupling potential.

The above RPA approximation may be considered as an extension to the
potential model. The extended potential model includes two channels
--- the ordinary valence quark channel and an extended sea quark
channel.  The valence quark channel which represents the contribution
of valence quarks is described by the x-amplitudes. The sea quark
channel representing the contribution of sea quarks is described by
the y-amplitudes. The ordinary quark potential $U$ interacts within
both individual channel. The new coupling potential $V$ couples the
two channels together.

The potentials $U$ and $V$ will be evaluated from the kernels in
action (\ref{A1}). Following ref.~\cite{Gromes77}, from the spinor
structure of Lorentz invariant, the kernels can be divided into five
types, i.e., scalar ($S$), pseudo-scalar ($P$), vector ($V$),
axial-vector ($A$), and tensor ($T$). We write
\begin{eqnarray}
\label{kernel}
K &=& S 1 \otimes 1 + P i\gamma^5 \otimes i\gamma^5
+ V \gamma^\mu \otimes \gamma_\mu 
+ A \gamma^\mu \gamma^5 \otimes \gamma_\mu \gamma^5
+ \frac12 T \sigma_{\mu\nu} \otimes \sigma^{\mu\nu}.
\nonumber \\
\end{eqnarray}
In the non-relativistic limit ${\bm k}, {\bm k}' \rightarrow 0$, only the
following terms survive in the potentials
\begin{eqnarray}
\label{U1}
U &=& - S + V + A {\bm \sigma}_1 \cdot {\bm \sigma}_2 
+ T {\bm \sigma}_1 \cdot {\bm \sigma}_2, \\
V &=& P - V {\bm \sigma}_1 \cdot {\bm \sigma}_2 - A 
- T {\bm \sigma}_1 \cdot {\bm \sigma}_2.
\end{eqnarray}
where ${\bm \sigma}$'s are the pauli spin matrices.

\section{Primary Application}

In addition to the Lorentz invariance of quark interaction kernels, in
the limit of vanishing current quark masses, the kernels must also
have the chiral symmetry of QCD. The general quark interaction which
is compatible with the chiral symmetry can be found in
ref.~\cite{Klimt90}.  Here we simply choose the form of effective
quark interaction $H_4$ as a vector interaction like the one gluon
exchange interaction
\begin{eqnarray}
\label{eff}
H_4 &=& 
\frac12 \int d^3x d^3y K({\bm x} - {\bm y})
\bar\Psi({\bm x}) \gamma^\mu {\bm\lambda}^c \Psi({\bm x})
\bar\Psi({\bm y}) \gamma_\mu {\bm\lambda}^c \Psi({\bm y}).
\end{eqnarray}

Following the quark potential model \cite{Isgur}, the effective quark
interaction kernel $K$ may be chosen as a confinement interaction plus
the one gluon exchange interaction
\begin{eqnarray}
K(r) &=&
- \frac3{16} b r + \frac14 \frac{\alpha_s}{r}.
\end{eqnarray}
Our choice of confinement interaction differs from ref.~\cite{Isgur}
in that ours is a vector $\gamma^\mu \otimes \gamma_\mu$ interaction.
This is because from the effective interaction kernel eq.~(\ref{kernel}),
according to ref.~\cite{Gromes77}, the effective $qq$ interaction is
\begin{eqnarray}
\label{U2}
U &=& S + V - A {\bm \sigma}_1 \cdot {\bm \sigma}_2 
+ T {\bm \sigma}_1 \cdot {\bm \sigma}_2.
\end{eqnarray}
Notice that scalar interaction $1 \otimes 1$ changes sign from the $q
\bar{q}$ (eq.~(\ref{U1})) to $qq$ (eq.~(\ref{U2})). This seems to us
that one can not get the confinement for both mesons and baryons
simutaneously from a unified interaction kernel.

To actually perform calculations with these potentials, we would like
to diagonalize the hamiltonian matrix in the harmonic-oscillator basis
following ref.~\cite{Isgur}. However, in this primary analysis, we
simply approximate the meson ground states using the ground state
wave function of harmonic-oscillator
\begin{eqnarray}
x({\bm k}, s_1, s_2) &=& X \Psi_{000}(k) 
\langle 1/2 s_1 1/2 s_2 \mid S M_S \rangle, \\
y({\bm k}, s_1, s_2) &=& Y \Psi_{000}(k) 
\langle 1/2 s_1 1/2 s_2 \mid S M_S \rangle,
\end{eqnarray}
where $S M_S$ is the spin of the meson, and 
\begin{eqnarray}
\Psi_{000}(k) &=& \frac1{(\sqrt{\pi}\beta)^{3/2}} 
e^{- k^2 / 2\beta^2}.
\end{eqnarray}

From the above simplification and the effective hamiltonian
eq.~(\ref{eff}), the RPA equations reduce to
\begin{eqnarray}
\left[ \begin{array}{cc} A & B \\ -B & -A \end{array} \right]
\left[ \begin{array}{c} X \\ Y \end{array} \right]
&=& m
\left[ \begin{array}{c} X \\ Y \end{array} \right],
\end{eqnarray}
where
\begin{eqnarray}
A&=& 2 \langle k^0 \rangle + \langle K \rangle, \\
B&=& - \langle K \rangle \langle {\bm \sigma}_1 \cdot {\bm \sigma}_2 \rangle,
\end{eqnarray}
and
\begin{eqnarray}
\langle {\bm \sigma}_1 \cdot {\bm \sigma}_2 \rangle &=& 2 S(S+1) - 3.
\end{eqnarray}
In above formula, $\langle \dots \rangle$ takes the average value.
In nonrelativistic limit ${\bm k} \rightarrow 0$, hence
$k^0 = \sqrt{{\bm k}^2 + m_c^2} \approx m_c$, so we have approximately
\begin{eqnarray}
A&\approx& 2 m_c + \langle K \rangle.
\end{eqnarray}

By introducing
\begin{eqnarray}
\alpha &=& X+Y, \\
\beta &=& X-Y,
\end{eqnarray}
the RPA equations become
\begin{eqnarray}
(A+B) \alpha &=& m \beta, \\
(A-B) \beta &=& m \alpha.
\end{eqnarray}

For a preliminary analysis, we will only consider two typical mesons ---
the well-known Goldstone boson $\pi$ and the vector meson $\rho$ which
is well described in the original quark potential model as $q\bar q$
valence quark pair.  For $\pi$, its orbital angular momentum $L=0$,
spin $S=0$, we have
\begin{eqnarray}
\label{PI1}
2(m_c + 2 \langle K \rangle) \alpha_\pi &=& m_\pi \beta_\pi, \\
\label{PI2}
2(m_c - \langle K \rangle) \beta_\pi &=& m_\pi \alpha_\pi.
\end{eqnarray}

The chiral symmetry gives a relation between constituent quark mass
$m_c$ and average value of effective quark interaction $\langle K
\rangle$. From MFA
\begin{eqnarray}
m_c &=& m_q + \Sigma,
\end{eqnarray}
where $m_q$ is the current quark mass, $\Sigma$ is the Hatree-Fock
quark self-energy. In the chiral limit $m_q \rightarrow 0$, the pion
will be an exact massless Goldstone boson, i.e., $m_\pi \rightarrow
0$. From the equations of pion, eqs.~(\ref{PI1}) and (\ref{PI2}),
we obtain in the chiral limit where $m_q=0$, $m_\pi=0$,
\begin{eqnarray}
\Sigma &=& \langle K \rangle.
\end{eqnarray}
Finally
\begin{eqnarray}
m_c &=& m_q + \langle K \rangle.
\end{eqnarray}

Back to equations of $\pi$ meson, we obtain
\begin{eqnarray}
m_\pi &=& 2 \sqrt{ m_q (3m_c -2m_q) }, \\
\frac{\alpha_\pi}{\beta_\pi} &=& \sqrt{\frac{m_q}{3m_c- 2m_q}}.
\end{eqnarray}

For $\rho$ meson, its $L=0$ and $S=1$, then 
\begin{eqnarray}
2m_c \alpha_\rho &=& m_\rho \beta_\rho, \\
2(m_c + \langle K \rangle ) \beta_\rho &=& m_\rho \alpha_\rho.
\end{eqnarray}
We obtain
\begin{eqnarray}
m_\rho &=& 2 \sqrt{ m_c (2m_c -m_q) }, \\
\frac{\alpha_\rho}{\beta_\rho} &=& \sqrt{\frac{2m_c -m_q}{m_c}}.
\end{eqnarray}

In the table~\ref{cal}, we give two sets of numerical calculations. In
the first calculation ($A$), the quark masses ($m_q$ and $m_c$) are
taken to be general accepted values. In the second calculation ($B$),
the quark masses are chosen to better fit the experimental data of
meson masses ($m_\pi$ and $m_\rho$). We believe that both sets of
results are reasonable as a primary calculation. The amplitude ratio
$Y_{\pi, \rho} / X_{\pi, \rho}$ determines the sea quark contribution
vs. the valence quark contribution \cite{Deng}. Since $X_\pi \approx
Y_\pi$ the sea quark contribution is essential to $\pi$ meson. For $\rho$
meson, $X_\rho \gg Y_\rho$, the sea quark contribution is small.

\begin{table}[ht]
\caption{Numerical calculations (all mass units are MeV).}
\label{cal}
\begin{ruledtabular}
\begin{tabular}{ccccccc}
& $m_q$ & $m_c$ & $m_\pi$ & $m_\rho$ & $X_\pi/Y_\pi$ & $X_\rho/Y_\rho$ \\ 
\hline
$A$ & 5.5 & 300 & 140 & 845 & $-$1.2 & 5.9 \\ \hline
$B$ & 6   & 274 & 139 & 771 & $-$1.2 & 5.9 \\
\end{tabular}
\end{ruledtabular}
\end{table}

Since $m_\pi \gg m_q$, we have approximately
\begin{eqnarray}
m_\pi &=& 2\sqrt3 \sqrt{ m_q m_c }, \\
\frac{\alpha_\pi}{\beta_\pi} &=& \sqrt{\frac{m_q}{3m_c}},\\
m_\rho &=& 2\sqrt2 m_c, \\
\frac{\alpha_\rho}{\beta_\rho} &=& \sqrt2 .
\end{eqnarray}
Then we find a mass relation
\begin{eqnarray}
m_\pi^2 &=& \sqrt{18} m_q m_\rho
\end{eqnarray}
The experimental data $m_\pi = 139$MeV, $m_\rho = 770$MeV combining
with the general accepted estimation $m_q = 5.5$MeV give
\begin{eqnarray}
\frac{m_\pi^2}{m_q m_\rho} &=& 4.6,
\end{eqnarray}
while $\sqrt{18}=4.2$.

\section{Summary and Discussion}

To take into account of the sea quark contribution to mesons more
properly, the original quark potential model is extended which now
includes a new sea quark channel similiar to the RPA approximation. A
new coupling potential is needed to couple the original valence quark
channel and the sea quark channel. Through an effective quark
interaction which is also chiral invariant as demanded by QCD, the new
potential is related to the conventional quark potential in original
quark potential model.  The constituent quark mass is related to the
quark potential also from the chiral symmetry.  The primary
calculation connected the masses of mesons to masses of quarks
(constituent and valence).  The sea quark contribution is important to
the $\pi$ meson which is a well-known Goldstone boson.  Other mesons,
such as vector $\rho$ meson, where the valence quark contribution is
dominant, can also be well described by the same extended quark
potential model.

Finally, we shall indicate that in this primary analysis, under the
non-relativistic approximation, many important features of quark
interaction such as the spin-orbit interaction, the spin-spin
interaction and tesor interaction are neglected. Those must be
considered in the complete calculation of meson spectroscopy in the
future.

\begin{acknowledgments}
This work is supported in part by China Natinal Nature Science Foundation.
\end{acknowledgments}

\end{document}